\documentclass[10pt, conference, letterpaper]{IEEEtran}
\IEEEoverridecommandlockouts
\usepackage{url}
\usepackage{amsmath,graphicx}
\usepackage{cite}
\usepackage{amsmath,amssymb,amsfonts}
\usepackage{algorithmic}
\usepackage{graphicx}
\usepackage{textcomp}
\usepackage{xcolor}
\usepackage{caption,subcaption}
\def\BibTeX{{\rm B\kern-.05em{\sc i\kern-.025em b}\kern-.08em
    T\kern-.1667em\lower.7ex\hbox{E}\kern-.125emX}}
    
\newcommand{\figVolunteer}{
  \begin{figure}
    \centering
    \includegraphics[width=1\columnwidth]{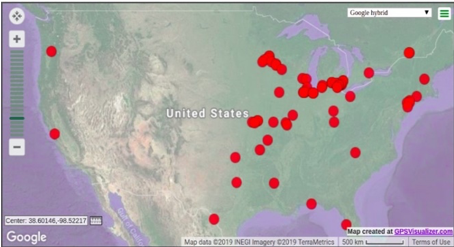}
    \caption{Volunteer sensor data log locations.}
    \label{fig:fig_volunteer}
  \end{figure}
}

\newcommand{\figNet}{
  \begin{figure}
    \centering
    \includegraphics[width=1\columnwidth]{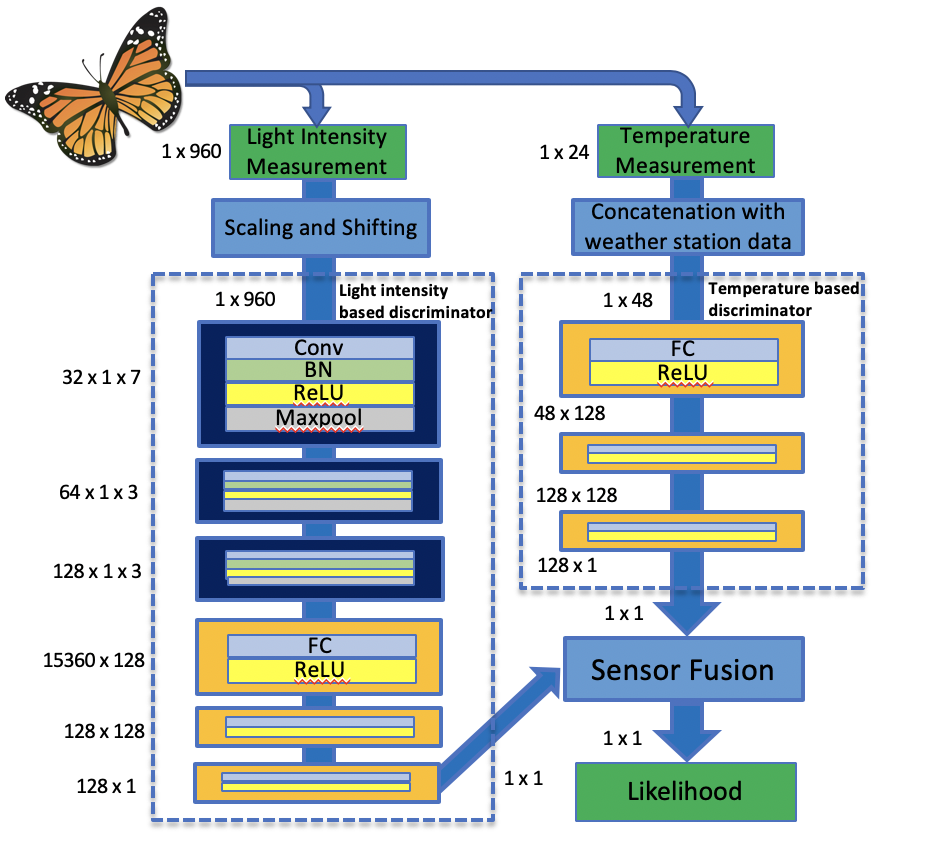}
    \caption{Light discriminator (left) and temperature discriminator (right) architecture.}
    \label{fig:fig_net}
  \end{figure}
}

\newcommand{\figLike}{
  \begin{figure}
    \centering
    \includegraphics[width=1\columnwidth]{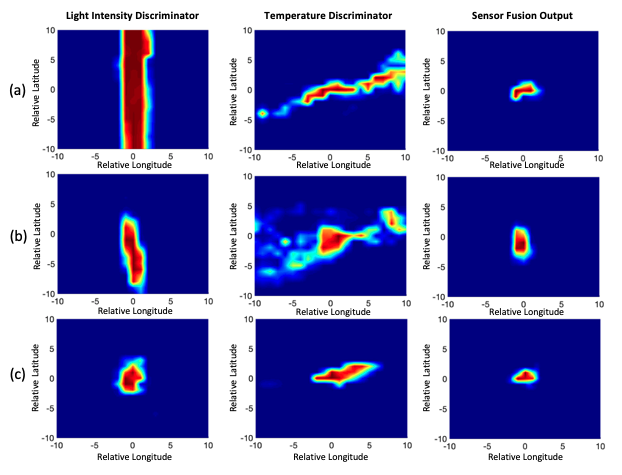}
    \caption{Example likelihood outputs for three different days; (a) 09/28, (b) 10/15, (c) 12/04.}
    \label{fig:fig_like}
  \end{figure}
}

\newcommand{\figResults}{
  \begin{figure}
    \centering
    \includegraphics[width=1\columnwidth]{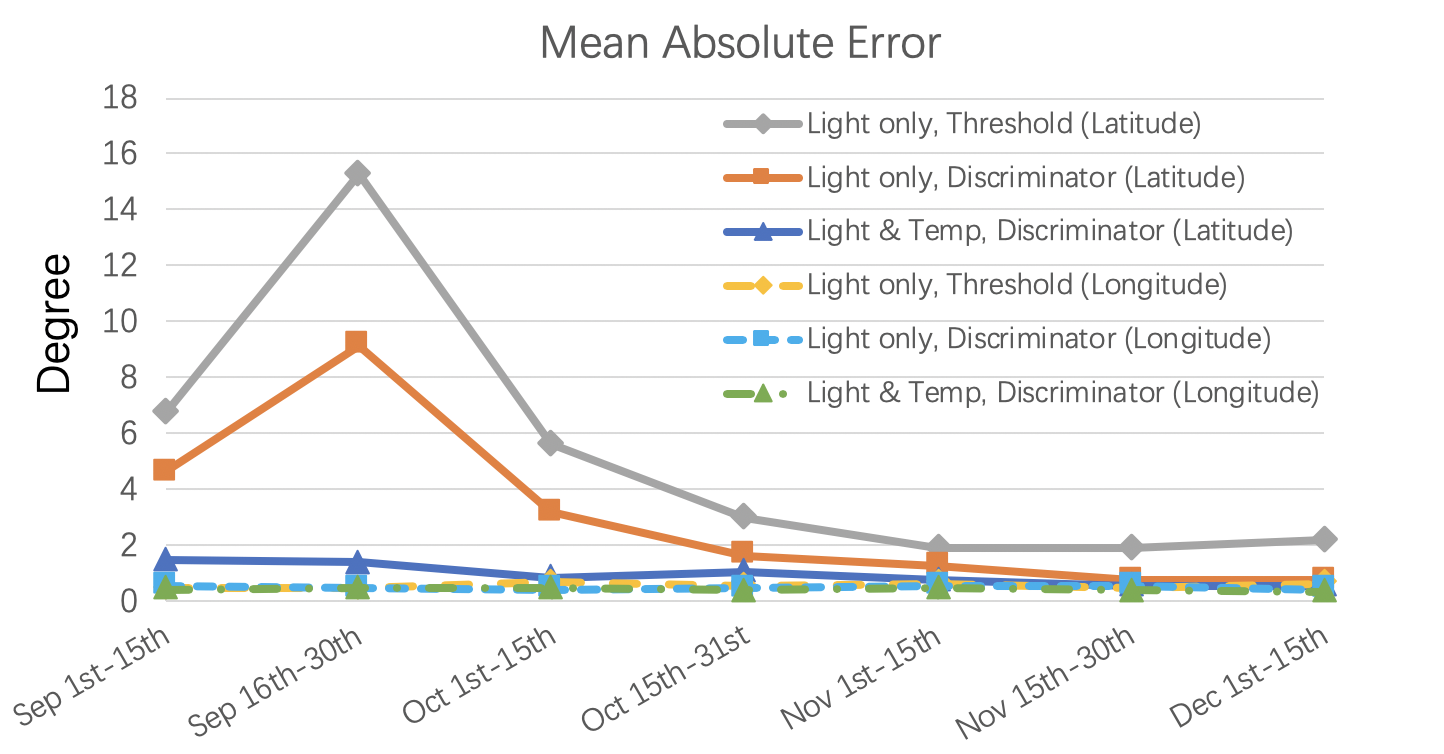}
    \caption{Mean absolute error of latitude and longitude evaluated biweekly.}
    \label{fig:fig_results}
  \end{figure}
}

\newcommand{\figLight}{
  \begin{figure}
    \centering
    \includegraphics[width=1\columnwidth]{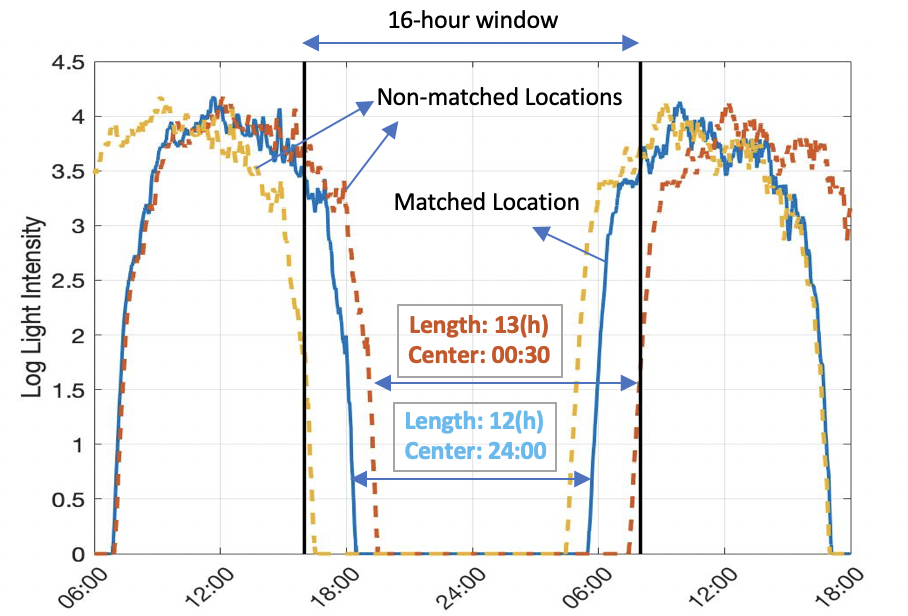}
    \caption{Example normalized light intensity curves.} 
    \label{fig:fig_light}
  \end{figure}
}

\title{Migrating Monarch Butterfly  Localization Using Multi-Sensor Fusion Neural Networks\\
\thanks{We thank the Monarch Butterfly Fund (MBF) for their support. We would also like to show our gratitude to all the volunteers who provided valuable light and temperature sensor measurements that greatly assisted this research. We could not have achieved these results without their help.  }
}

\author{\IEEEauthorblockN{Mingyu Yang$^\star$, Roger Hsiao$^\star$, Gordy Carichner$^\star$, Katherine Ernst$^\star$,\\ Jaechan Lim$^\star$, Delbert A. Green II$^\star$, Inhee Lee$^\dagger$, David Blaauw$^\star$, and Hun-Seok Kim$^\star$} 

\IEEEauthorblockA{$^\star$ University of Michigan, Ann Arbor, MI, USA
    \& $^\dagger$ University of Pittsburgh, Pittsburgh, PA, USA}
}

\begin{document}

\maketitle

\begin{abstract}
Details of Monarch butterfly migration from the U.S. to Mexico remain a mystery due to lack of a proper localization technology to accurately localize and track butterfly migration. In this paper, we propose a deep learning based butterfly localization algorithm that can estimate a butterfly's daily location by analyzing a light and temperature sensor data log continuously obtained from an ultra-low power, mm-scale sensor attached to the butterfly. To train and test the proposed neural network based multi-sensor fusion localization algorithm, we collected over 1500 days of real world sensor measurement data with 82 volunteers all over the U.S. The proposed algorithm exhibits a mean absolute error of $<1.5^{\circ}$ in latitude and $<0.5^{\circ}$ in longitude Earth coordinate, satisfying our target goal for the Monarch butterfly migration study. 
\end{abstract}
\section{Introduction}
\label{sec:intro}

Each fall, millions of Monarch butterflies across central and eastern U.S. and southern Canada migrate up to 2,500 miles to overwinter in the same location in central Mexico. In spring these migrants mate and remigrate northwards to repopulate their northern breeding territory over 3 -- 5 partially overlapping generations. Because no migrant Monarch lives long enough to make a return trip to the overwintering site, this navigational task cannot be learned and must be a genetically-encoded spatiotemporal program. 

At present, only the largest animal migrators can be tracked continuously for significant portions of their migratory journey (e.g., \cite{thorup2017resource}). Monarch butterflies, as small insects, cannot be tracked using the same strategy due to the weight and power constraints for mounted devices. A recent effort tracked Monarchs and green darner dragonflies up to hundreds of miles using the Motus Wildlife Tracking System \cite{knight2019radio}. While a substantial advance, this method has several limitations, such as unacceptable tracker weight, excessive power consumption, and very limited coverage of the Monarch migration territory.   

While the global positioning system (GPS) is the most conventional method for determining locations, the smallest commercial GPS solution \cite{Pin} has a total weight of 1 gram and size of 5cm, which is vastly too heavy and large for butterflies to carry. As an alternate to GPS, we propose to use daylight and temperature logging that are able to be integrated into the Michigan Micro Mote (M3) platform \cite{lee2012modular}\cite{jung2014ultra}\cite{jeong2014fully}, which has a potential of weight less than 50 mg and a size of $8\times8\times2.6\text{mm}^3$. Its duty-cycled operation is sustained by solar energy harvesting and it supports 50m distance wireless readout at the Monarch overwintering site. A new M3 platform customized for Monarch butterfly mounting is currently under development. 

This paper introduces a deep neural network based Monarch butterfly localization algorithm that utilizes the light intensity and temperature measurement data logged on the M3 platform. The proposed algorithm will be performed offline to analyze the migration trajectory when the log data is wirelessly retrieved from the butterfly at the overwintering site. To train and evaluate the proposed neural networks before deploying the final M3 system, we have conducted a sensor data measurement campaign with 82 volunteers across the U.S. to record solar light intensity and temperature using commercial HOBO sensors \cite{HOBO} as an emulator of the final M3 platform for the duration of Monarch fall migration in 2018. 

\vspace{-0.2cm}
\section{Related Work}

Light intensity based localization has been applied to tracking marine animals that remain submerged and out of the reach of GPS \cite{musyl2003vertical}\cite{wilson2005movements}\cite{wilson2006movements}\cite{nielsen2007state}. Prior attempts tried to explicitly estimate the sunrise and sunset time from recorded light intensity curve. Then longitude and latitude coordinates are determined based on the estimated day center and day length respectively using standard astronomical equations \cite{smith1986determining}\cite{welch1999assessment}\cite{hill2001geolocation}\cite{ekstrom2004advance}. However, it has the fundamental limitation of large latitude ambiguity around the equinox days (September 22 and March 20) when the day length is globally the same regardless of latitude. A second main challenge is the significant local light intensity variation due to weather and terrain factors that an ideal sunlight intensity model is unable to capture. To mitigate these issues, prior works \cite{lam2010incorporating}\cite{nielsen2006improving} augmented sea-surface temperature to improve the accuracy of tracking large sea animals for GPS-failing conditions (i.e., underseas). These approaches first construct a sea-surface temperature contour map based on satellite data and then localize the sensor by finding the position on the map where the temperature matches to the sensor reading. Although it is effective for sea animal tracking, the same method is not directly applicable to on-/above-ground butterfly localization where the temperature is significantly affected by the local terrain and weather.
\vspace{-0.2cm}
\section{Proposed Method}\label{sec:prop}
\subsection{Overview}
As a dynamic system, the butterfly daily localization problem can be described using a state-space framework, whose system model and measurement model are expressed by $\mathbf{x}_D =\mathbf{f}(\mathbf{x}_{D-1}) + \mathbf{w}_{D}$ and $[\mathbf{l}_{D}, \mathbf{t}_{D}]^{\top} = \mathbf{g}(\mathbf{x}_D) + \mathbf{n}_D$
where $D$ is the day index (with a unit of a day), $\mathbf{x}_D$ denotes the state vector that represents the latitude and longitude coordinate of the butterfly on the day $D$, and $\mathbf{l}_D$ and $\mathbf{t}_D$ denotes the discrete sequence of light intensity and temperature sensor data measured on the day $D$, respectively. $\mathbf{w}_D$ and $\mathbf{n}_D$ represents the sequence of process noise and observation noise, respectively. $\mathbf{f}$ is a function modeling the transition of state vectors and $\mathbf{g}$ is the observation function relating the state vector to measurements. In the Bayesian approach for dynamic state estimation, the goal is to construct the posterior $p(\mathbf{x}_D|\mathbf{l}_{1:D}, \mathbf{t}_{1:D})$ and then compute MMSE from posterior mean or MAP from posterior maximum. These posteriors can be formulated by prior $p(\mathbf{x}_0)$, likelihood $p(\mathbf{l}_D, \mathbf{t}_D |\mathbf{x}_D)$ and state transition probability $p(\mathbf{x}_D|\mathbf{x}_{D-1})$. 

In this paper, we mainly focus on the likelihood $p(\mathbf{l}_D, \mathbf{t}_D |\mathbf{x}_D)$ to localize the butterfly's daily position. As it is practically infeasible to find an exact expression of $\mathbf{f}$ and $\mathbf{g}$ for Monarch butterfly migration, we rely on deep neural networks to learn the implicit observation model based on real world data. Then, we treat the output of the neural network as the estimation of likelihood. Because light intensity and temperature measurements have different properties, we apply two distinct neural networks to learn their likelihoods separately. That is, $p(\mathbf{l}_D|\Tilde{\mathbf{x}}) \approx \Phi_l(\mathbf{l}_D,\Tilde{\mathbf{x}})$ and $p(\mathbf{t}_D|\Tilde{\mathbf{x}}) \approx \Phi_D(\mathbf{t}_D,\Tilde{\mathbf{x}})$ where $\Phi_l$ and $\Phi_t$ denote the two neural networks, $\Tilde{\mathbf{x}}$ denotes an arbitrary location. Then, with a simplifying assumption that light intensity and temperature measurement are conditionally independent given the state vector, we have $p(\mathbf{l}_D, \mathbf{t}_D|\Tilde{\mathbf{x}}) = p(\mathbf{l}_D|\Tilde{\mathbf{{x}}})p(\mathbf{t}_D|\Tilde{\mathbf{x}}) \approx \Phi_l(\mathbf{l}_D,\Tilde{\mathbf{x}})\Phi_t(\mathbf{t}_D,\Tilde{\mathbf{x}})$. We call $\Phi_l$ the light intensity discriminator and $\Phi_t$ the temperature discriminator. 

\subsection{Light Intensity Discriminator}
\figLight
In this section, we propose a light discriminator network to estimate $p(\mathbf{l}_D|\Tilde{\mathbf{x}})$. For the network input, we first define a reshape function $r$: $\hat{\mathbf{l}}_{D} = r(\mathbf{l}_{D}, \Tilde{\mathbf{x}}, D)$
where $\hat{\mathbf{l}}_{D}$ is the \textit{normalized} light intensity data obtained by shifting and resampling the original $\mathbf{l}_{D}$ based on the coordinate state $\Tilde{\mathbf{x}}$ and the date information $D$ so that the night center is located at the center and the length of the night is scaled to 12 hours as depicted in Fig. \ref{fig:fig_light}. The reason we normalize the light intensity curve based on the night center and length instead of the day center and length is because Monarch butterflies are known to rest without changing the location during the night. 

The input to the light discriminator network is $\hat{\mathbf{l}}_{D}$ reshaped from the observation $\mathbf{l}_{D}$ based on a state candidate $\Tilde{\mathbf{x}}$ given the measurement date information $D$. The neural network is trained to discriminate (i.e., binary classification) whether $\Tilde{\mathbf{x}}$ matches to the true measurement location or not by observing the shape of $\hat{\mathbf{l}}_{D}$ normalized based on $\Tilde{\mathbf{x}}$. To generate the training dataset for this discriminator network, we use both matched and unmatched pairs of $(\mathbf{l}_{D}, \Tilde{\mathbf{x}})$. The final activation funtion of the discriminator network is the softmax. Hence the output $\Phi_l(\tilde{\mathbf{l}}_D, \Tilde{\mathbf{x}})$ can be interpreted as the likelihood probability $p(\mathbf{l}_D|\Tilde{\mathbf{x}})$. 


The length of $\hat{\mathbf{l}}_{D}$ as the input of the discriminator is set to $\pm$8 hours around the night center. Longer window length increases the complexity of the neural network unnecessarily without improving the discriminator accuracy as it mostly rely on the night center position, night length, and the light intensity data shape near the sunrise and sunset time for binary classification. Reducing the input length is also beneficial to avoid overfitting when the training dataset size is limited. 

\subsection{Temperature Discriminator}

The temperature discriminator network is designed to estimate $p(\mathbf{t}|\Tilde{\mathbf{x}})$. However, unlike the light intensity data, it does not have any dependency to the longitude coordinate of the measurement location and it is also significantly dependent on the local weather. Thus, we train the discriminator to compare the two inputs; the temperature measurement data from the sensor and the weather station measurement data at a particular location on the same day. It produces the binary classification result; matched or unmatched depending on whether the location of the weather station is closest to $\Tilde{\mathbf{x}}$ or not. The final softmax function of the discriminator network quantifies the temperature data pattern similarity between the sensor and weather station data at a particular location $\Tilde{\mathbf{x}}$, estimating  $\Phi_t(\mathbf{t}_D,\Tilde{\mathbf{x}}) \approx p(\mathbf{t}_D|\Tilde{\mathbf{x}})$. 

\section{Experiments}
\label{sec:exp}
\subsection{Hardware \& Data Collection}
\figVolunteer

For data collection, we use HOBO sensors \cite{HOBO} as an emulator of the final M3 platform to record the light intensity and temperature data. To collect the real world data, we disseminated HOBO sensors to 82 volunteers in the U.S. This volunteer data contain 1604 valid night measurements with a time resolution of 10s for light intensity and 15s for temperature from 1st September to 19th December in 2018. The volunteer sensor placement locations are shown in Fig. \ref{fig:fig_volunteer}. We access the night temperature weather station data with time resolution of 1 hour using WeatherBit API \cite{WEATHER}. 
The time offset and resampling factor for the light intensity reshaping function $\hat{\mathbf{l}}_{D} = r(\mathbf{l}_{D}, \Tilde{\mathbf{x}}, D)$ are obtained by the astronomical equation MATLAB function \cite{SUNRISE} which calculates the sunrise and sunset time (which are converted to the night center time and night length) for a given coordinate $\Tilde{\mathbf{x}}$ on the day $D$.

\subsection{Data Processing and Preparation}

1604 days of valid sensor measurements data are divided into 1300 training and 304 testing data. The light intensity data are down-sampled to 1 minute resolution and then converted to log scale. Since the WeatherBit weather station temperature data has time resolution of 1 hour, we also down-sample temperature data to 1 hour interval to match the sampling rate. 




For each of the 1300 training light intensity data ${\mathbf{l}}_{D}$, we prepare one matched pair $(\hat{\mathbf{l}}_{D}, \Tilde{\mathbf{x}})$ and 24 unmatched pairs by applying random night center and night length offset $\Delta t$ in the range of $2 \text{ minutes} \leq |\Delta t| \leq 2$ hours. We end up with a training set of size 32500 in which 1300 entries are labeled Class 1 (match) and 31200 entries are Class 0 (non-match) for binary classification. 
A similar process is applied to generate the training dataset for the temperature discriminator. 
For each of 1300 temperature measurements ${\mathbf{t}}_{D}$ training data, we prepare one matched pair using the weather station data at the nearest location to label it Class 1 (matched). In addition, we create 15 unmatched pairs with the Class 0 label for each sensor measurement ${\mathbf{t}}_{D}$ using the weather station data randomly selected from 15 different locations in the range of $[-20,20]$ degrees in both latitude and longitude around the ground-truth location. When there is no weather station data in the vicinity of a random position, we simply treat it as an outlier without adding it to the training dataset. This approach leads to 17198 Class 0 data and 1300 Class 1 data in total.   

\subsection{Network structure}
\figNet
The network structures for the proposed discriminators are shown in Fig. \ref{fig:fig_net}. The light intensity discriminator contains 3 convolution layers (conv - batch normalization - ReLU - max pooling) and 3 fully connected layers. To avoid overfitting, a dropout layer with $p=0.25$ is added after the first fully connected layer. The size of each layer is specified in Fig. \ref{fig:fig_net}. The temperature discriminator network shown on the right in Fig. \ref{fig:fig_net} only contains three fully connected layer and a dropout layer with $p=0.25$ placed after the first fully connected layer. Since the size of Class 0 dataset is much larger than the size of Class 1 for both networks, we adopt a weighted sampling technique that samples the dataset unevenly so that the two classes are equally distributed for each batch. 

\subsection{Results}
The proposed neural network based likelihood estimation ($\Phi_l(\mathbf{l}_D,\Tilde{\mathbf{x}})$ and $\Phi_t(t_D,\Tilde{\mathbf{x}})$) is performed by collecting the neural network output for each test data ${\mathbf{l}}_{D}$ and ${\mathbf{t}}_{D}$ evaluated at various coordinates $\Tilde{\mathbf{x}}$ in a grid surrounding the ground-truth sensor location with a range of $[-10,10]$ degrees in latitude and longitude. 
The $\Tilde{\mathbf{x}}$ grid resolution for the initial (coarse) likelihood evaluation of $\Phi_l(\mathbf{l}_D,\Tilde{\mathbf{x}})$ and $\Phi_t(\mathbf{t}_D,\Tilde{\mathbf{x}})$ is 1 degree for both longitude and latitude. The spatial resolution of the likelihood estimation is refined to 0.1 degree by upsampling (and interpolating) the coarse evaluation results. 

\figLike

Three example likelihood estimations on three different days are shown in Fig. \ref{fig:fig_like} where red and blue color corresponds to high and low likelihood, respectively. The plots on the left, center, and right column show the (interpolated) neural network output $\Phi_l(\mathbf{l}_D,\Tilde{\mathbf{x}})$, $\Phi_t(\mathbf{t}_D,\Tilde{\mathbf{x}})$, and the product $\Phi_l(\mathbf{l}_D,\Tilde{\mathbf{x}})\Phi_t(\mathbf{t}_D,\Tilde{\mathbf{x}})$, respectively, based on a randomly selected sensor data instance $(\mathbf{l}_D, \mathbf{t}_D, \mathbf{x})$ while the ground-truth sensor location ${\mathbf{x}_D}$ is shifted to (0,0) for plotting. The sensor measurement data on the row (a), (b), and (c) were collected on the date of Sep. 28,  Oct. 15, and Dec. 4, respectively. We observed that the light discriminator neural network mostly relies on the night length information to estimate the latitude while it uses the night center time information to estimate the longitude. Thus, when the night length is globally the same regardless of the coordinate around the equinox day (row (a) on Sep. 28), the light discriminator network fails to estimate the longitude and it produces a pattern of $\Phi_l(\mathbf{l}_D,\Tilde{\mathbf{x}})$ spread out along the latitude as shown on the top left of Fig. \ref{fig:fig_like}. 

While the light discriminator has high ambiguity in latitude around the equinox and maintains low ambiguity in longitude, the opposite is true for the temperature discriminator as temperature varies significantly along latitude but less so along longitude as shown in Fig. \ref{fig:fig_like} middle column. Therefore, light and temperature discriminator networks uniquely complement each other, resulting in significant accuracy improvement. When two neural network outputs are multiplied, it provides more reliable results to estimate the likelihood $\Phi_l(\mathbf{l}_D,\Tilde{\mathbf{x}})\Phi_t(\mathbf{t}_D,\Tilde{\mathbf{x}}) \approx p(\mathbf{l}_D, t_D|\Tilde{\mathbf{x}})$ as shown in Fig. \ref{fig:fig_like} right column. In general, the output has smaller error in December (Fig. \ref{fig:fig_like} row (c)) when both the light and temperature discriminators work reliably due to significant night length, night center and temperature variations across latitude and longitude.

\figResults

Finally, we evaluate the  localization accuracy using the test dataset and we compare the proposed approach to a baseline where the sunrise, sunset time and night length are estimated by comparing the light intensity to a threshold calibrated for the minimum error. 
The average longitude and latitude localization errors for different time intervals are shown in Fig. \ref{fig:fig_results}. All methods have similar performance in longitude estimation exhibiting less than $0.5^{\circ}$ average absolute error for all periods. For latitude, our light discriminator significantly outperforms the baseline method for all periods. By combining the likelihood estimation from the light and temperature networks, the average error in latitude reduces dramatically from $9^{\circ}$ to $1.5^{\circ}$ around the fall equinox.

\section{Conclusion}

We present a neural network based butterfly localization algorithm that learns the observation model implicitly. The proposed method is applicable to ultra-low power and ultra-small light and temperature sensors that can be attached to Monarch butterflies without impeding their migration. The maximum likelihood localization confirms that neural networks can learn implicit observation models to outperform hand-craft models. Testing results exhibit 1-degree error of latitude/longitude, which is sufficient to study Monarch migration. We will continue collecting more volunteer measurements to improve the robustness of the neural networks. 

\bibliographystyle{IEEEbib}
\bibliography{refs}

\end{document}